\documentclass[showpacs,reprint,amsmath,amssymb,pra]{revtex4-1}

\usepackage{graphicx}
\usepackage{dcolumn}
\usepackage{bm}
\usepackage{subfigure}
\begin{document}
\title{Nonlocal  Gap Solitons in $\mathcal{PT}$-symmetric periodic potential with Defocussing Nonlinearity}
 \author{Chandroth P. Jisha$^{1*}$, Alessandro Alberucci $^2$, Valeriy A. Brazhnyi,$^{1}$ and Gaetano Assanto$^{2}$}
\affiliation{
$^1$Centro de F\'{\i}sica do Porto, Faculdade de Ci\^encias, Universidade do Porto, R. Campo Alegre 687, Porto 4169-007, Portugal\\
$^2$Nonlinear Optics and OptoElectronics Lab (NooEL), Via della Vasca Navale 84, 00146 Rome, Italy \\ $*$Corresponding author: cpjisha@gmail.com
}

\begin{abstract}
Existence and stability of $\mathcal{PT}$-symmetric gap solitons in a periodic structure with defocussing nonlocal nonlinearity are studied both theoretically and numerically. We find that, for any degree of nonlocality, gap solitons are always unstable in the presence of an imaginary potential. The instability manifests itself as a lateral drift of solitons due to an unbalanced particle flux. We also demonstrate that the perturbation growth rate is proportional to the amount of gain (loss), thus predicting the observability of stable gap solitons for small imaginary potentials.
\end{abstract}

\maketitle

\section{Introduction}\label{sec:intro}
Systems with non-Hermitian Hamiltonian  possessing a real eigenvalue spectrum gained interest after the early work by Bender et al., \cite{Bender:1998}, who showed that the condition of Hermiticity to obtain real eigenvalues can be replaced if the system  satisfies an additional condition of parity-time invariance in the form  $\bold{PTH} = \bold{HPT}$, with the additional constraint that {\bf H} and {\bf PT} share a common set of eigenvectors, with {\bf P} and {\bf T}  parity and time reversal operators, respectively. Moreover, it was demonstrated that quantum systems encompassing a potential such that $V(x)=V^*(-x)$ satisfy the $\mathcal{P} \mathcal{T}$ symmetry.  In fact, the eigenvalues remain purely real in a given subset of system parameters: a phase transition towards complex eigenvalues connected with a spontaneous symmetry breaking can be observed \cite{Bender:2005,Guo:2009}. Such systems are appealing  from both fundamental and practical perspectives, as evident from the  vast literature in diverse 
areas, including quantum mechanics, integrated optics, plasmonics and metamaterials among others \cite{Bender:1999, Ramezani:2012, Kottos:2012, Benisty:2011, Bender:2013, Castaldi:2013, Lazarides:2013}. 

In optics, where the attention paid to $\mathcal{P}\mathcal{T}$ theory has been higher, the interest relies mainly on the fact that the Maxwell equations can be often recast as a Schr{\"o}dinger equation; the biggest advantage is the ease in realizing complex Hamiltonians both in the spatial \cite{Ruter:2010} and in the temporal domains \cite{Regensburger:2012, Regensburger:2013} and, consequently, the accessible experimental demonstration of the theoretical findings. Singular light properties such as nonreciprocal photon propagation have been demonstrated, leading to new  technological achievements like the realization of an all-optical diode in Silicon \cite{Feng:2011}. The role of nonlinearity has been addressed as well, with the prediction of $\mathcal{PT}$-symmetric solitons \cite{Musslimani:2008}. In optical systems, $\mathcal{PT}$-symmetric potentials can be obtained by suitable tailoring of the real and imaginary parts of the refractive indices, such that the real part is a symmetric function with 
position and the imaginary part encompassing gain and loss terms is anti-symmetric in space (including metal insertions \cite{Guo:2009,Feng:2011} and pumped system \cite{Ruter:2010, Hang:2013});  in matter wave systems $\mathcal{PT}$-symmetric potentials can be achieved by suitable gain and loss mechanisms introduced through coupling  with external reservoirs.

On the other hand, in the last decade attention has been paid to waveguide arrays and the formation of discrete solitons \cite{Lederer:2008, Fratalocchi:2004}, including spatial gap solitons \cite{Conti:2000,Mandelik:2004}, that is, self-localized nonlinear waves with a propagation constant within the linear bandgap of the periodic structure. Light propagation through $\mathcal{PT}$-symmetric periodic potentials, including exotic phenomena such as $\mathcal{PT}$-symmetry breaking, nonreciprocal behavior and double refraction were investigated from the beginning, both in the linear and nonlinear regimes \cite{Makris:2008,*Makris:2010,*Makris:2011, Musslimani:2008, Dmitriev:2010}. The existence of $\mathcal{PT}$-symmetric solitons and breathers was also discussed in the context of couplers and array geometries \cite{Alexeeva:2012, *Barashenkov:2012, *Barashenkov:2013}. The existence and stability of two-dimensional gap solitons in $\mathcal{PT}$ linear lattices with a 
local Kerr response  was discussed in~\cite{Zeng:2012}.

While stable gap solitons can exist in $\mathcal{PT}$-symmetric linear periodic potentials in nonlinear defocusing local media, the role of  nonlocality on the stability and mobility of localized solutions is still an open issue. Recently, studies were reported on solitons in $\mathcal{PT}$-symmetric periodic potentials in nonlocal self-focusing media  \cite{Li:2012} and in the presence of linear defects \cite{Hu:2012}, as well as on nonlocal bright solitons in defocusing media \cite{Nixon:2012}, including geometries with localized $\mathcal{PT}$-symmetric potentials \cite{Shi:2012}. Nevertheless, the effect of nonlocality on gap solitons in $\mathcal{PT}$-symmetric periodic potentials with self-defocusing was never addressed to date.

In the present work, we study the existence and stability of $\mathcal{PT}$-symmetric gap solitons in defocusing nonlocal media in the presence of a periodic potential and with a diffusive-like nonlinear response, such that exhibited by, e.g., nematic liquid crystals \cite{Fratalocchi:2004,Piccardi:2011_1}, thermo-optic materials \cite{Conti:2009} and atomic vapors \cite{Suter:1993}. 
While stationary modes are obtained only on-site (i.e., corresponding to potential minima) due to lack of parity symmetry for the overall system, we demonstrate that gap solitons in $\mathcal{PT}$-potentials are always unstable for any degree of nonlocality,  generally undergoing oscillatory instabilities (OI). Multi-component and multi-parameter systems as well as dissipative systems are commonly associated with OI, the latter originally discussed in the context of parametrically driven  damped Kerr media \cite{Barashenkov:1991} and later extended to the generalized Thirring model \cite{Rossi:1998, Barashenkov:1998} and other systems \cite{Yuanyao:2009, Johansson:1999}. It is to be noted that the existence of OI in Hermitian systems results in the eventual decay or blow-up of the stationary solution \cite{Johansson:1999, Rossi:1998}. Here we demonstrate that the existence of OI in $\mathcal{PT}$-symmetric systems leads to unidirectional energy transfer from one lattice site to the other, 
specifically from the gain region towards the loss region. 

\section{Defocusing nonlocal gap solitons in the bandgap region}\label{sec:ngs}

We consider a generic field $\Psi$ describing either the particle distribution for matter waves or the electric field for light (electromagnetic) waves. We further assume that the material is linearly inhomogeneous and can be modeled  by a linear periodic potential $V(x)$ with a non-vanishing imaginary part \cite{Makris:2008}. Finally, we study a nonlocal nonlinear medium with a diffusive response. The evolution of the field $\Psi$ versus propagation  $z$ (time for matter waves, propagation distance for light) obeys the following system

\begin{eqnarray}
\label{eq:GPE}
&& i\frac{\partial \Psi}{\partial z}=-\frac{1}{2}\frac{\partial^2 \Psi}{\partial x^2}+V(x)\Psi+V_\textrm{NL}(|\Psi|^2)\Psi,\\
&& V_\textrm{NL}- \sigma \frac{\partial^2 V_\textrm{NL}}{\partial x^2}=|\Psi|^2, \label{eq:Yukawa}
\end{eqnarray}
where $x$ is the transverse coordinate and $V_\textrm{NL}$ is the nonlinear portion of the complete potential $V + V_\textrm{NL}$. Equation \eqref{eq:GPE} is the well-known Gross-Pitaevskii equation (i.e. nonlinear Schr\"odinger equation in optics), whereas Eq. \eqref{eq:Yukawa} is in the form of a screened Poisson or Yukawa equation. The parameter $\sigma$ (defined as a strictly positive quantity) determines the range of nonlocality: a large $\sigma$ corresponds to a highly nonlocal response \cite{Alberucci:2007}; in particular, the width $d$ of the Green function $G(x)$ of Eq. \eqref{eq:Yukawa} is proportional to $\sqrt{\sigma}$, with $G(x)=-1/(2\sqrt{\sigma}) \exp{(-|x|/\sqrt{\sigma})}$. In writing Eq. \eqref{eq:Yukawa} we assumed a self-defocusing Kerr medium as $V_\mathrm{NL}$ is proportional to the intensity $|\Psi|^2$:  a bell-shaped $\Psi$ gives rise to a   repulsive potential. For the potential to be $\mathcal{PT}$-symmetric, a necessary (but not sufficient) condition is 
that $V(x) = V^*(-x)$ \cite{Ruter:2010}: hereby we will make the ansatz $V(x)= V_r(x) + iV_i(x) 
=V_r\sin^2(x) - i V_i \sin(2x)$, setting to $\pi$ the period of the linear potential without any loss of generality. 

We look for propagation invariant solutions of Eqs. (\ref{eq:GPE})-(\ref{eq:Yukawa}) in the form $\Psi(x,z)=\phi(x)\mathrm{exp}(-i \mu z)$, being $-\mu$ the propagation constant for light waves or $\mu$ the chemical potential for matter waves. The substitution into (\ref{eq:GPE})-(\ref{eq:Yukawa}) provides the nonlinear eigenvalue problem

\begin{eqnarray}
&&\mu\phi=-\frac{1}{2}\phi_{xx}+ [V_r\mathrm{sin}^2(x) - iV_i \sin(2x)] \phi+ V_\mathrm{NL} \phi,\label{eq:phi}\\
&& V_\mathrm{NL} -\sigma\frac{\partial^2 V_\mathrm{NL}}{\partial x^2}=|\phi|^2. \label{eq:sol_VNL}
\end{eqnarray}

\subsection{Linear eigenmodes}
\label{sec:linear}

The linearized version of Eq. (\ref{eq:phi}) (i.e., with $V_\mathrm{NL}=0$) supports Bloch waves of the form $\phi_K(x) = u_K(x)\exp(i\,K\,x)$ \cite{Kittel:1995} where $u_K(x+\pi) = u_K(x)$, i.e. $u$ is a $\pi$-periodic function of $x$.

\begin{figure}
\centering
\includegraphics[width=0.5\textwidth]{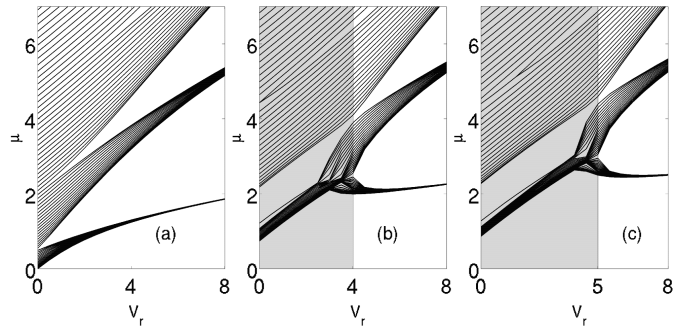}
\caption{\label{Fig:band} Real part of the eigenvalue $\mu$ versus $V_r$ with the imaginary part $V_i$ of the $\mathcal{PT}$-symmetric potential taking the value (a) $0$, (b)  $2$ and (c)  $2.5$, respectively. The energy spectrum has complex eigenvalues in the shaded region.}
\end{figure}

Figure ~\ref{Fig:band} shows the behavior of the linear eigenvalues $\mu$ versus the real potential $V_r$ for  purely real ($V_i=0$) and for $\mathcal{PT}$-symmetric ($V_i\neq 0$) potentials \cite{Musslimani:2008}. For $V_i=0$ the number and size of the bandgaps increase with $V_r$ due to the stronger confinement in each lattice site. For $V_i\neq 0$ the two lowest bands merge for small $V_r$ (inside the shaded region), corresponding to the break-up of the $\mathcal{P}\mathcal{T}$-symmetry: the eigenvalues $\mu$ belonging to this band are complex. When $V_r$ overcomes a threshold, dependent on $V_i$ and explicitly given by $V_r=2V_i$, the two lowest bands split, the $\mathcal{P}\mathcal{T}$-symmetry is fulfilled and the eigenvalues become purely real \cite{Musslimani:2008,Makris:2011}, corresponding to the unshaded region in Fig. \ref{Fig:band}. 

\begin{figure}
\centering
\includegraphics[width=0.45\textwidth, height = 6cm]{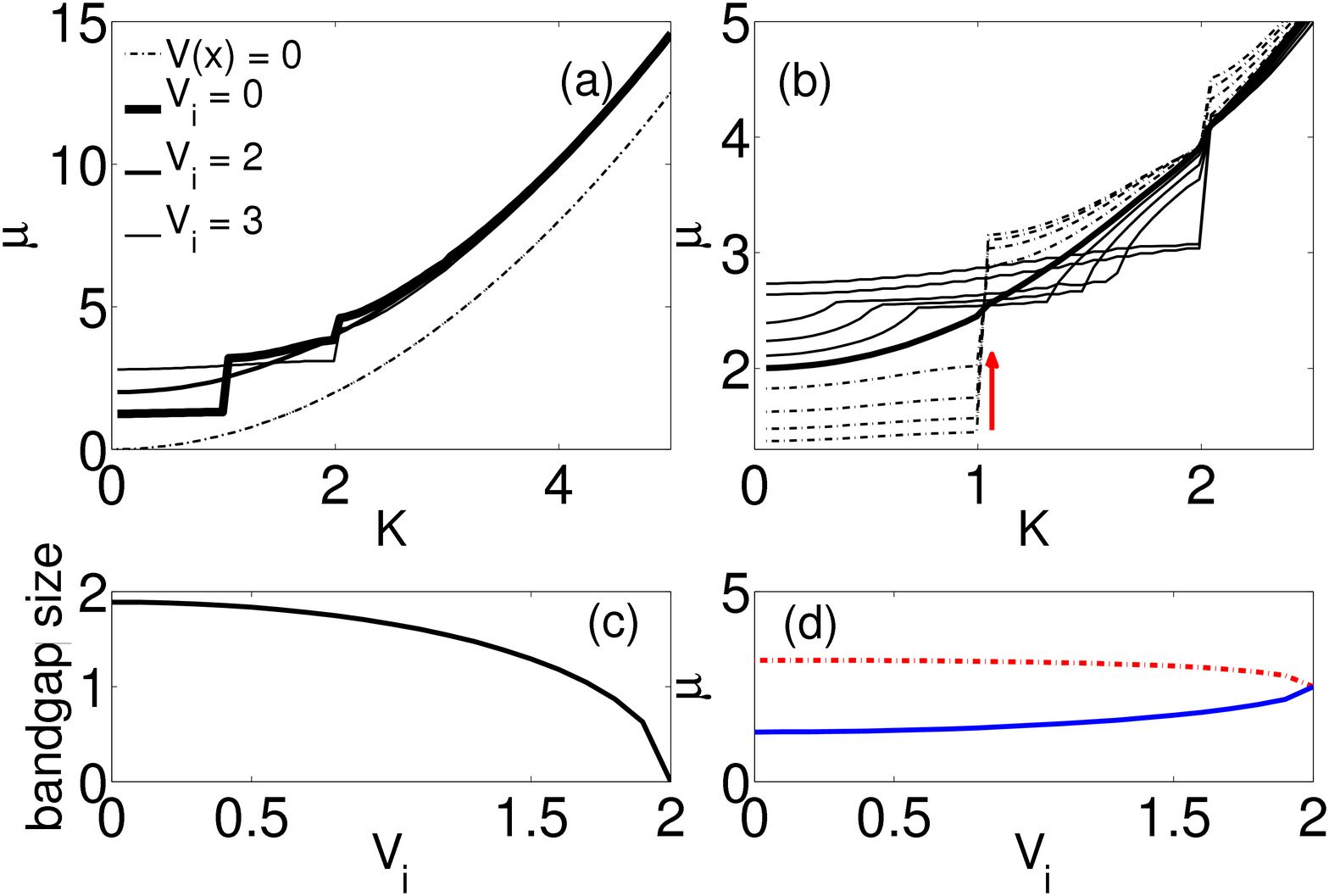}
\caption{\label{Fig:disp} Dispersion relation for $V_r = 4$. (a) Real part of $\mu$ for $V_i$ varying from 0 to 3 (thicker to thinner lines). The bottom line shows the free particle case. Note the curve for $V_i=2$, where the bandgaps disappear and the dispersion relation follows $\mu = K^2/2 + V_c$: this is the $\mathcal{PT}$-symmetry breaking point. Below this critical point all the eigenvalues are real. (b) Enlargement of the dispersion relation. The arrow indicates the direction of increasing $V_i$. Dotted lines: $V_i< 2$; thin lines: $V_i>2$. (c) Bandgap  size and  (d) $\mu$ values at the lower (solid blue line) and upper (dotted red line) band-edges versus $V_i$. }
\end{figure}

The eigenvalue spectrum for the free particle case, i.e. in the absence of any potential, follows $\mu= K^2/2$. As well known, for $V_i =0$ the system spectrum exhibits gaps; in particular, for $V_r=4$, there are two finite bandgaps for $\mu<8$ (thickest line in Fig. \ref{Fig:disp}(a)). Figure \ref{Fig:disp} graphs the dependence of the dispersion relation on the imaginary part $V_i$ of the potential when $V_r$ is arbitrarily fixed to 4, together with the free particle case (lowest curve). Starting from $V_i=0$ and then increasing $V_i$ (see Fig. \ref{Fig:disp} (c-d)), the eigenvalue spectrum remains real ($\mathcal{P}\mathcal{T}$-symmetry is conserved) until $V_i = V_r/2 = V_c$, where the potential reduces to $V(x) = V_r [1 - \exp(2ix)] /2$: thus, for $V_i=V_c$ the energy bandgaps disappear, resulting in a spectrum equivalent to that of a free particle, but shifted above by $V_c$, i.e.  $\mu = K^2/2 + V_c$ \cite{Makris:2011, Nixon:2012}. Conversely, when $V_i>V_c$ (broken $\mathcal{P}\mathcal{
T}$-symmetry) the spectrum becomes complex; furthermore, there is no energy bandgap in the first Brillouin zone, as the first bandgap shifts to the second Brillouin zone [see Fig. \ref{Fig:disp} (b)]. Such symmetry breaking in the spectrum 
was discussed in detail in \cite{Musslimani:2008, Midya:2010, Makris:2010}.

For $V_i = V_c$ and neglecting the nonlinearity, the Floquet mode $u_K(x)$ obeys (up to the end of the section we will consider a potential of period $T$ in order to generalize our result)

\begin{eqnarray}
\mu u_K&=& -\frac{1}{2}\frac{\partial^2 u_K}{\partial x^2}+ \frac{ V_r + K^2}{2} u_K \nonumber \\&-& i K \frac{\partial u_K}{\partial x} - \frac{V_r e^{\frac{2\pi i x}{T}}} {2}u_K \label{eq:vc}
\end{eqnarray}

with $T=\pi$ in our case.
Equation \eqref{eq:vc} can be solved by expressing $u_K(x)$ in its Fourier series as $u_K(x)=\sum_{n=-\infty}^\infty a_n(K) e^{\frac{2\pi in x}{T}}$. After defining $G=K+2\pi n /T$, a direct substitution into Eq. \eqref{eq:vc} provides the recursive relation

\begin{equation}
  \left(\frac{V_r + G^2}{2} - \mu \right) a_n = \frac{V_r}{2} a_{n-1}. \label{eq:recursive_formula}
\end{equation}

By looking at Eq. \eqref{eq:recursive_formula} it is apparent that, if $\left(\frac{V_r + G^2}{2} - \mu \right)\neq 0$, when one of the coefficients $a_n$ vanishes then all $a_n$ go to zero, leading to the trivial solution $u_K=0$. Hence, non-trivial Bloch modes exist only if $G^2=2\mu-V_r$, with $G^2$ spanning all the positive half of the real axis; thus, the eigenvalue $\mu$ has to satisfy the necessary condition

\begin{equation} \label{eq:existence_condition}
  \mu > \frac{V_r}{2}=V_c.
\end{equation}

For a fixed pair $(\mu , K)$ we have a non-zero $u_K(x)$ if there is a given $\overline{n}(\mu,K)$ such that $0.5 \left( K + 2 \pi \overline{n} / T \right)^2 = \mu - V_r/2$; then the expansion coefficients $a_n(\mu,K)$ are zero for $n<\overline{n}-1$, whereas for $n>\overline{n}-1$ they can be computed via Eq. \eqref{eq:recursive_formula}. Summarizing, the system shows a continuous spectrum with a cutoff at $\mu = V_c $, in agreement with our numerics \cite{Nixon:2012}.		\\
For the sake of simplicity and without loss of generality, hereafter we  take $V_r =4$: Fig. \ref{Fig:band} (b) shows that the transition from a real to a complex spectrum occurs for $V_i=2$.

\begin{figure}
\centering
\includegraphics[width=0.5\textwidth]{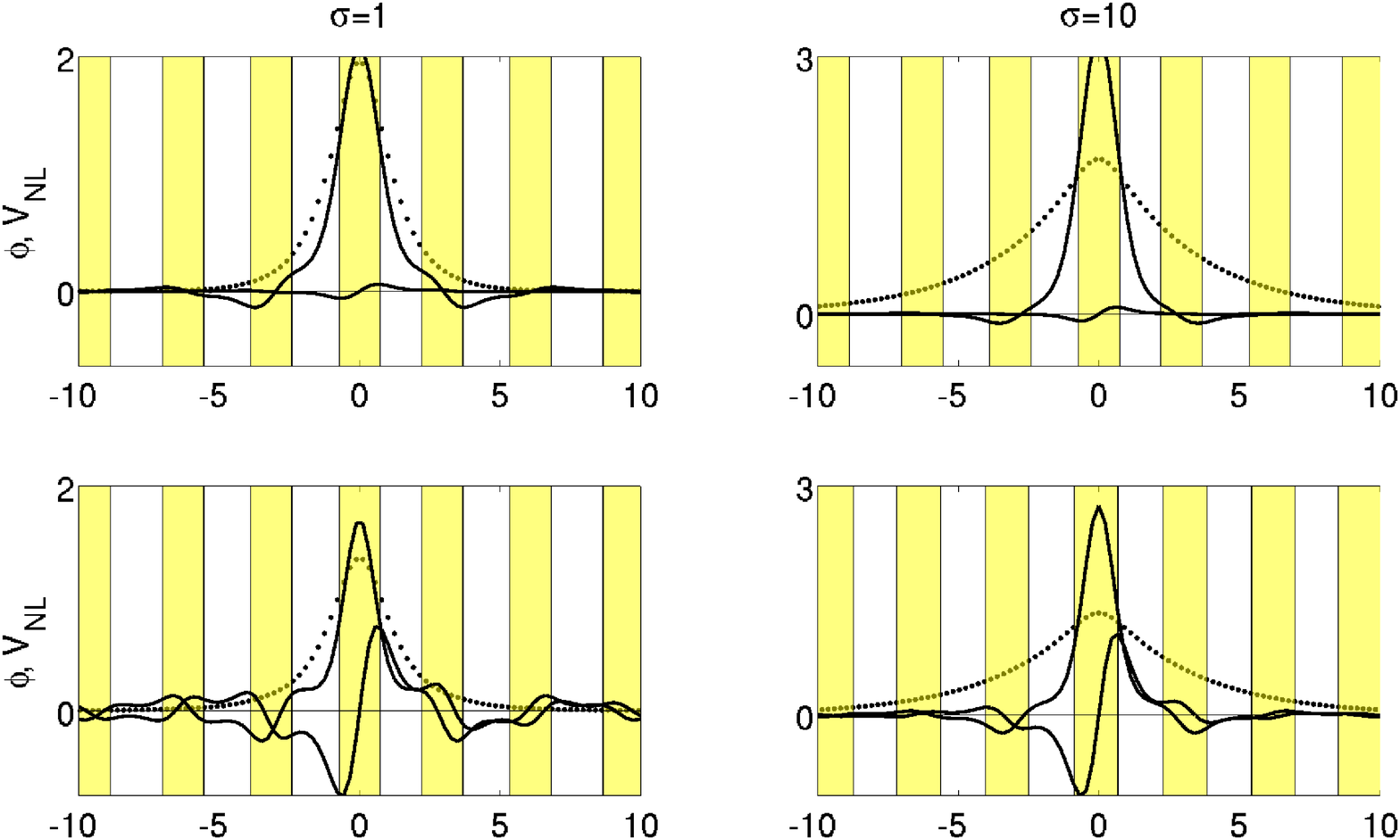}
\caption{\label{Fig:profile} Gap soliton profiles for $\mu=3$ versus $x$. The symmetric and anti-symmetric curves correspond to the real and  imaginary parts, respectively. The dotted lines show the corresponding nonlinear potential $V_\textrm{NL}$ for two values of the nonlocality  $\sigma=1$ (left) and $\sigma=10$ (right). Each row corresponds to $V_i = 0.1$ and $1.5$, respectively. The yellow/white boxes sketche the qualitative trend of the linear real potential $V_r(x)$ along $x$, with mid-points in the yellow and white regions corresponding to minima and maxima of the real potential, respectively; otherwise stated, the yellow regions correspond to the guiding channels. }
\end{figure}

\subsection{Nonlinear case} 
 
We now concentrate on calculating the form assumed by shape-preserving wavepackets in the presence of nonlinearity. As is well known, gap solitons exist only in the linear bandgaps (hence the name): we will look for fundamental $\mathcal{PT}$-symmetric gap solitons embedded in the first linear bandgap by choosing $V_i< V_c$ to avoid symmetry breaking. We also note that, due to the presence of the odd imaginary potential $V_i$, off-site gap solitons do not exist in our case \cite{Jisha:2010}. Typical profiles of on-site gap solitons, obtained using a pseudo-spectral technique based on Chebyshev polynomials, are shown in Fig. \ref{Fig:profile} for various degrees of nonlocality  and for $\mu$ close to the upper band-edge. We set a numerical window  much wider than the nonlocal length $d$ to avoid spurious effects from the boundaries \cite{Alberucci:2007,Efremidis:2008}. 
Analogously to the case of purely real lattices \cite{Jisha:2010,Xu:2005}, the real component of the fundamental gap soliton is mainly localized in a given guide, with tails extending towards adjacent guides and shaped so that the linear modes in these guides are excited in phase opposition with respect to the main lobe \cite{Lederer:2008}. Moreover, regardless of $V_i$, the larger the nonlocality is the lower the tails are (compare left and right columns in Fig. \ref{Fig:profile}), similar to the case $V_i=0$ \cite{Jisha:2010}. A physical explanation of this behavior relies on  coupled mode theory: to exist, gap solitons require out-of-phase excitation of neighboring guides, with the nonlinear response providing the needed difference in propagation constants (energies for matter waves) between core and side channels. In the local Kerr  case, a strong excitation is needed in the adjacent sites to reach the necessary nonlinear phase modulation, whereas in the nonlocal case the nonlinear perturbation induced 
by 
the mode in the core guide, spreading outwards owing to the finite size of the Green function $G(x)$, provides the required modulation. 

At variance with the case $V_i=0$, gap solitons have a symmetric real part  and an anti-symmetric imaginary part in order to fulfill power (number of particles) conservation. In fact, in the presence of a complex potential the particle conservation for the imaginary Schr\"odinger equation (that is, containing a complex potential) reads

\begin{equation}  \label{eq:particle_conservation}
  \nabla \cdot j = -\frac{\partial \rho}{\partial z} + 2 V_i(x){\rho},
\end{equation}
where $\rho=|\phi|^2$ is the wave intensity and the particle flux $j$ is given by $j=\frac{1}{2i} \left(\psi^* \frac{\partial \psi}{\partial x} - \psi \frac{\partial \psi^*}{\partial x} \right)$. Setting $\phi=\sqrt{\rho}e^{i\chi(x)}$, the flux reads $j=\rho \frac{\partial \chi}{\partial x}$. Setting $\partial\rho/\partial z=0$, Eq. \eqref{eq:particle_conservation} yields

\begin{equation} \label{eq:flux}
  \frac{\partial }{\partial x} \left(\rho \frac{\partial \chi}{\partial x} \right)= 2V_i(x) {\rho}.
\end{equation}

\begin{figure}
\centering
\includegraphics[width=0.5\textwidth]{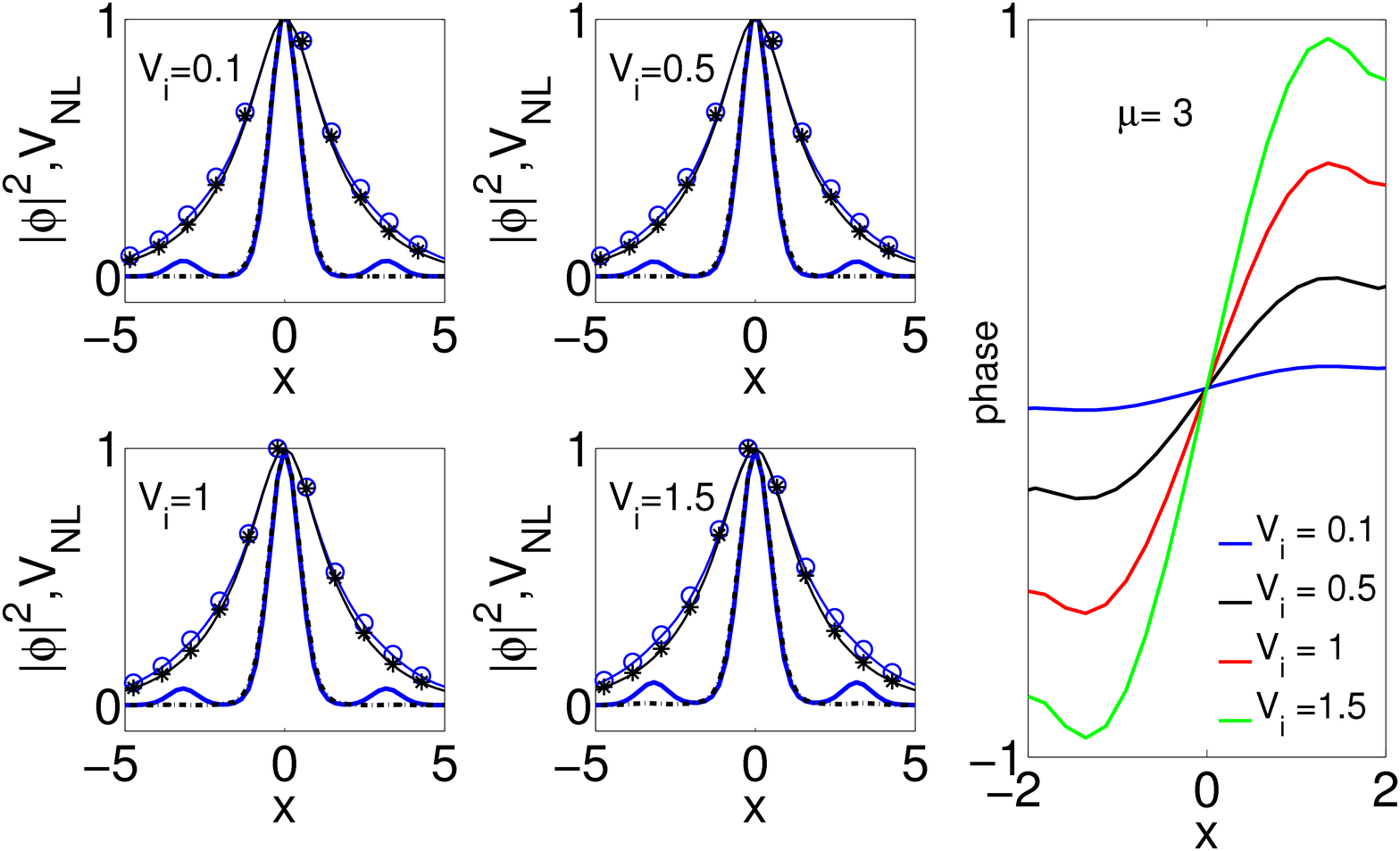}
\caption{\label{Fig:profiled} Normalized stationary profiles $|\phi|^2$ (lines) and corresponding nonlinear potential $V_\textrm{NL}$ (symbols) 
for various $V_i$ and $\sigma=10$ when $\mu= 1.4$, $1.45$, $1.6$ and $1.9$ (lowest $\mu$ for each $V_i$, $\circ$, solid line) and $\mu = 3$  ($*$, dotted line). An increase in $\mu$ reduces the soliton tails and makes the solution  more and more confined in the channel where initially excited. The last panel shows the phase profile of the solutions when $\mu=3$; at each $x$ the absolute phase increases with $V_i$.  }
\end{figure}

According to Eq. \eqref{eq:flux}, an even $\rho$ corresponds to an odd $\chi$, that is, an even real part and an odd imaginary part, respectively. Physically, particles are created within the gain regions and then they diffuse (transversely) towards the loss regions in order to keep  the overall (i.e., integrated along $x$) particle number constant. Noteworthy, the flux $j$ is an even function, that is, a unidirectional flow takes place \cite{Musslimani:2008}. Equation \eqref{eq:flux} also states that the larger $V_i$ is the larger the anti-symmetric part is (see the rightmost panel in Fig. \ref{Fig:profiled}), due to the increase in the transverse flux of particles necessary to compensate the inhomogeneous gain/loss. The flux $j$  corresponds in optics to the transverse component of the Poynting vector, see Ref. \cite{Musslimani:2008}.

Figure \ref{Fig:profiled} elucidates the dependence of the soliton tails on the propagation constant $\mu$, i.e., on the soliton power. For values of $\mu$ close to the lower band-edge [see Fig. \ref{Fig:disp}(d)], the solitons have pronounced tails (solid lines without symbols in Fig. \ref{Fig:profiled}), with the tails diminishing as the eigenvalue $\mu$ approaches the upper band-edge (dotted lines without symbols in Fig. \ref{Fig:profiled}), such dynamics being fully analogous to what happens in a purely real potential \cite{Jisha:2010}. The phases associated with these solutions are illustrated in the last panel of Fig. \ref{Fig:profiled}: the phase has  hyperbolic tangent profile in the central region and deviates from it when overlapping with the adjacent guides. 

\begin{figure}
\centering
\includegraphics[width=0.5\textwidth]{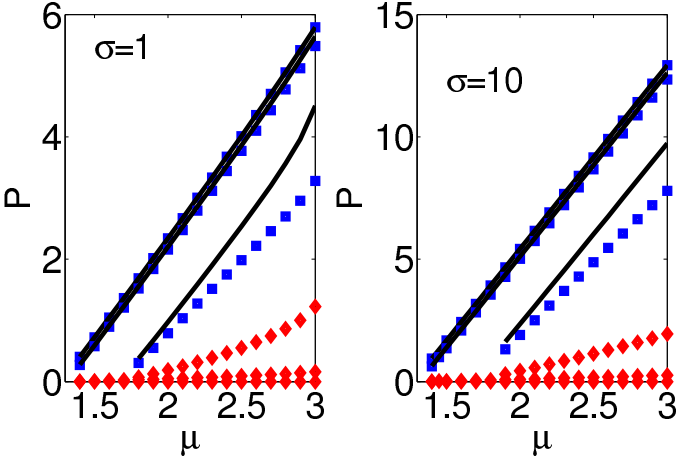}
\caption{\label{Fig:Pmu} Power $P$ vs $\mu$ for $\sigma=1$ and $\sigma=10$, respectively. The black solid curves represent the total power $P$, the blue squares ($\Box$) correspond to $P_r$ (top to bottom) and the red diamonds ($\diamond$) to $P_i$ for $V_i = 0$, $0.5$ and $1.5$, from bottom to top, respectively. Fundamental gap solitons are considered here.}
\end{figure}

Figure \ref{Fig:Pmu} plots the soliton power  (number of particles for matter waves), defined as $ P = \int |\phi|^2 dx$. First, the soliton power increases with the nonlinear eigenvalue $\mu$, consistently with the self-defocusing nature of the nonlinear response. The power $P_{r,i}=\int |\phi_{r,i}|^2 dx$ carried by the real and  the imaginary parts of the solution is also graphed. With increasing magnitude of the complex potential, the power carried by the imaginary part also increases, in turn affecting the stability of the solutions as we will demonstrate in the following. We also note that the power carried by the soliton increases with the nonlocality $\sigma$ due to the lower nonlinear effect (see the form of the Green function $G$ and Ref. \cite{Jisha:2010}), with a ratio $P_r/P_i$ roughly proportional to $V_i$ but slightly dependent on $\sigma$. 

\section{Variational approach}

Before studying the stability of the solutions, we adopt a variational analysis to look  for analytical solutions and verify how well this semi-analytical approach  predicts the solutions of a $\mathcal{PT}$-symmetric system.  We believe this is the first time that variational analysis is applied to optical $\mathcal{PT}$-symmetric systems.
The semi-analytical results using a variational approximation \cite{Cerda:1998,Jisha:2010, Jisha:2005} are obtained from the Lagrangian of the system Eq. (\ref{eq:GPE}) for solutions of the form $\Psi(x,z)=\phi(x,z)\mathrm{exp}(-i \mu z)$;   for the conservative part the Lagrangian reads

\begin{eqnarray}
 L _{C}&=& \frac{i}{2}(\phi_z^* \phi - \phi_z \phi^*) - \mu|\phi|^2+\frac{1}{2}|\phi_x|^2+V_r\mathrm{sin}^2(x)|\phi|^2 \nonumber \\ & + & V_\mathrm{NL} |\phi|^2-\frac{\sigma}{2}\left(\frac{\partial V_\mathrm{NL}}{\partial x} \right)^2-\frac{V_\mathrm{NL}^2}{2}. \label{eq:Lagn}
\end{eqnarray}

As Fig. \ref{Fig:profiled} shows, the width of the intensity profile does not appreciably vary for increasing $\mu$, the main portion of the wavepacket being confined in a single channel. At the same time, the exact soliton profiles have tails which, for $\mu$ close to the upper band, become negligibly small. This allows us to choose very general trial functions of the form 

\begin{eqnarray}
 &&\phi=A \exp\left(-\frac{x^2}{ w_b^2}\right) \exp\left[ i \theta(z) f(x)\right], \label{eq:trial_field} \\ 
 &&V_\mathrm{NL} = A_\mathrm{NL} \exp\left(-\frac{x^2}{ w_\mathrm{NL}^2}\right), \label{eq:trial_index}
\end{eqnarray}
where $A$ corresponds to the field amplitude, $\theta$ is the amplitude of the phase profile and  $f(x)$ is its spatial distribution along $x$; $A_\mathrm{NL}$ is the amplitude of the nonlinear potential and, finally, $w_b$ and $w_\mathrm{NL}$ are the widths of the soliton and of the nonlinear perturbation $V_\mathrm{NL}$, respectively. Given that the real part of the solution is symmetric and that the imaginary part is anti-symmetric, we assume $f(x)$ to be an odd function of $x$. The nonlocal nonlinear potential well is parabolic-like in the proximity to the soliton and proportional to the Green function of Eq. \eqref{eq:Yukawa} away from it (i.e., $\exp(-|x|/d)$); for simplicity we take a Gaussian profile. The phase factor is accounted for in the $\theta$ term and, as shown below, can predict the observed behavior of the solutions to a reasonable extent. 

The standard variational approach for systems with dissipative terms can be  modified as \cite{Cerda:1998,Ankiewicz:2007}

\begin{equation} \label{lagvar}
\frac{d}{dz}\left(\frac{\partial \langle L _{C}\rangle}{\partial
\beta_{z}}\right)- \frac{\partial \langle L _{C}\rangle}{\partial
\beta}=2 \text{Re}{\int_{-\infty}^{\infty}}Q\frac{\partial
\phi^*}{\partial \beta}dx,
\end{equation}

where $\beta$ stands for all the parameters free to vary in our variational computation $(A$, $ A_\mathrm{NL}$, $P$, $w_b$, $w_\mathrm{NL}$, $\theta)$;  we also defined $Q = i V_i \sin(2x) \phi$ and $\langle L _{C}\rangle = \int _{-\infty} ^\infty L_C dx$. \\
Using Eqs. (\ref{eq:trial_field}-\ref{eq:trial_index}), the reduced Lagrangian is 

\begin{eqnarray}
\langle& L _{C}&\rangle = \sqrt{\frac{2}{\pi}} \frac{P}{w_b} \theta_z \int_{-\infty}^{\infty}  f(x) e^{-\frac{2x^2}{w_b^2}} dx - \mu P + \frac{P}{2 w_b^2} \nonumber \\ \nonumber
&+&  \sqrt{\frac{2}{\pi}} \frac{P}{2 w_b} \theta^2 \int_{-\infty}^{\infty} f_x^2 e^{-\frac{2x^2}{w_b^2}} dx + \frac{V_r P (1 - e^{-\frac{w_b^2}{2}})}{2}  \\ 
&+& \frac{\sqrt{2} A_\mathrm{NL} P w_\mathrm{NL}}{(2w_\mathrm{NL}^2 + w_b^2)^{1/2}} - \frac{ \sqrt{\pi} \sigma A_\mathrm{NL}^2}{2 \sqrt{2} w_\mathrm{NL}} - \frac{\sqrt{\pi} w_\mathrm{NL} A_\mathrm{NL}^2}{2\sqrt{2} }. \label{eq:variational_1}
\end{eqnarray}

The first term on the RHS of Eq. \eqref{eq:variational_1} vanishes owing to the anti-symmetry of $f(x)$. Using Eq. \eqref{lagvar}, variational equations are obtained for each variable parameter. The RHS of Eq. \eqref{lagvar} is nonzero only for $\beta = \theta$ because $f(x)$ is the only odd function; the substitution of Eq. \eqref{eq:variational_1} into \eqref{lagvar} provides

\begin{equation}
\theta = -2 V_i \frac{ \text{Re}{ \left[\int_{-\infty}^{\infty} \sin(2x) f(x) e^{-\frac{2x^2}{w_b^2}} dx \right]}}{  \int_{-\infty}^{\infty} f_x^2 e^{-\frac{2x^2}{w_b^2}} dx} . \label{eq:theta} 
\end{equation}

\begin{figure}
\centering
\includegraphics[width=0.5\textwidth]{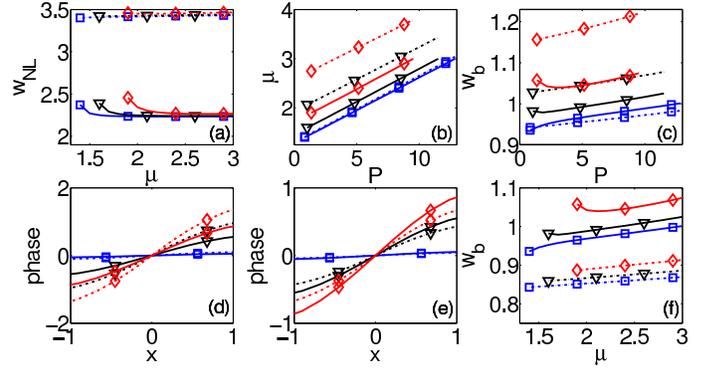}
\caption{\label{Fig:var} Graphs of (a) $w_\mathrm{NL}$ versus $\mu$, (b) $\mu$  versus $P$, (c)  $w_b$  versus $P$  and  (d) phase profile, calculated using the variational approach. (e) Phase profile  and (f) $w_b$ calculated using a simplified analytical approach (see text). In all  panels $V_i = 0.1$ ($\Box$), $V_i = 1$ ($\triangledown$), $V_i = 1.5$ ($\diamond$)  and $\sigma=10$. Solid lines correspond to numerics and dotted lines  are variational (analytical) results.  }
\end{figure}

Equation \eqref{eq:theta} clearly shows that the phase associated with the solution depends on the imaginary part of the refractive index $V_i$; it vanishes when the potential is purely real, eventually providing a flat-phase soliton.\\
Similarly, the variation of Eq. \eqref{lagvar} with respect to $A_\mathrm{NL}$ and $w_\mathrm{NL}$ yields 

\begin{equation} \label{eq:wNL_variational}
w_\mathrm{NL}^2 = \frac{w_b^2 + 2 \sigma + \sqrt{(w_b^2 + 2 \sigma)^2 +24\sigma w_b^2}}{4},  
\end{equation}
in full analogy with a conservative system \cite{Jisha:2010}. Figure \ref{Fig:var} (a) compares Eq.  \eqref{eq:wNL_variational} with numerical solutions: the two approaches are in good agreement (flat behavior versus $\mu$), except for a constant factor due to the different shapes of the nonlinear perturbation. Clearly, the variational approach is not able to model the broadening of $V_\mathrm{NL}$ near the band-edge due to the presence of non-negligible tails in the actual soliton.\\
The effects of the complex potential are apparent in the expression for $w_b$ 

\begin{eqnarray} \label{wbvar}
&&\frac{1}{w_b^4}  =   \frac{V_r}{2}e^{-\frac{w_b^2}{2}} +\sqrt{\frac{2}{ 4 \pi w_b^2}} \theta^2  \frac{\partial}{\partial w_b} \left(\frac{1}{w_b}\int_{-\infty}^{\infty} f_x^2 e^{-\frac{2x^2}{w_b^2}}dx\right)  \nonumber \\
&-&   \frac{2 w_\mathrm{NL}^3 P}{(2w_\mathrm{NL}^2 + w_b^2)^2(\sigma+w_\mathrm{NL}^2)}, 
\end{eqnarray}

and $\mu$

\begin{eqnarray} \label{wbvarmu}
&&\mu =  \frac{1}{2 w_b^2} + \frac{V_r}{2}(1-e^{-\frac{w_b^2}{2}}) +\sqrt{\frac{2}{ 4 \pi w_b^2}} \theta^2 \int_{-\infty}^{\infty} f_x^2 e^{-\frac{2x^2}{w_b^2}}dx  \nonumber \\
&+&   \frac{2 w_\mathrm{NL}^3 P}{(2w_\mathrm{NL}^2 + w_b^2)(\sigma+w_\mathrm{NL}^2)}.
\end{eqnarray}

Figure \ref{Fig:profiled} suggests a hyperbolic tangent profile for the transverse phase distribution, i.e. $f(x)$. Therefore we take $f(x) = \mathrm{tanh}(x)$ and plot the beam width using Eq. \eqref{wbvar} for various $P$ and $V_i$  in Fig. \ref{Fig:var}(c). The propagation constant $\mu$ is evaluated using Eq. \eqref{wbvarmu}, as well, and is compared with numerical results in  Fig. \ref{Fig:var}(b). The dependence on soliton power $P$ and imaginary potential $V_i$ match in both cases: in fact, $w_b$ increases with power owing to self-defocusing, whereas the soliton broadens with $V_i$ because of the larger flux of particle to accommodate. Fig. \ref{Fig:var} (d) plots the numerical and variational phase profiles for various $V_i$, with differences ascribable to the erroneous evaluation of $w_b$ in the variational method: the substitution of $w_b$  from numerics into Eq. \eqref{eq:theta} 
provides a nearly perfect agreement. 

In the highly nonlocal limit we can get another set of simplified analytical formulae for $w_b$ and $\mu$. Following Ref. \cite{Snyder:1997} we can set $V_\mathrm{NL}\approx V_\mathrm{NL}^{(0)}+V_\mathrm{NL}^{(2)}x^2$, with $V_\mathrm{NL}^{(0)}=\int{ G(-x') |\phi(x')|^2 dx'}$ and $V_\mathrm{NL}^{(2)}=\left. \left(V_\mathrm{NL}^{(0)} - |\phi(0)|^2 \right)\right/(2\sigma)$. Thus, Eq. \eqref{eq:phi} provides 

\begin{eqnarray} \label{wbsimp1}
&&\frac{2}{w_b^4}  =   V_r - \theta^2  +  V_\mathrm{NL}^{(2)},
\end{eqnarray}

\begin{eqnarray} \label{wbsimp2}
&&\mu   =  \frac{1}{w_b^2} + \frac{\theta^2}{2} + V_\mathrm{NL}^{(0)},
\end{eqnarray}
and 
\begin{equation}\label{thetasimp}
 \theta = \frac{2V_i}{1+ \frac{2}{w_b^2}};
\end{equation}
the corresponding results  are plotted in Fig. \ref{Fig:var}(e)-(f). The observed discrepancy in phase profile can be attributed to the fact that $w_b$ calculated using this approach is shifted by a small constant value (Fig. \ref{Fig:var} (f)). Using $w_b$ from numerics in Eq. \eqref{thetasimp} yields a perfect agreement with the numerically evaluated phase profile.

\section{Stability and dynamical evolution}

\subsection{Linear stability analysis (LSA)}
\label{sec:LSA}

\begin{figure*}
\centering
\includegraphics[width=0.8\textwidth]{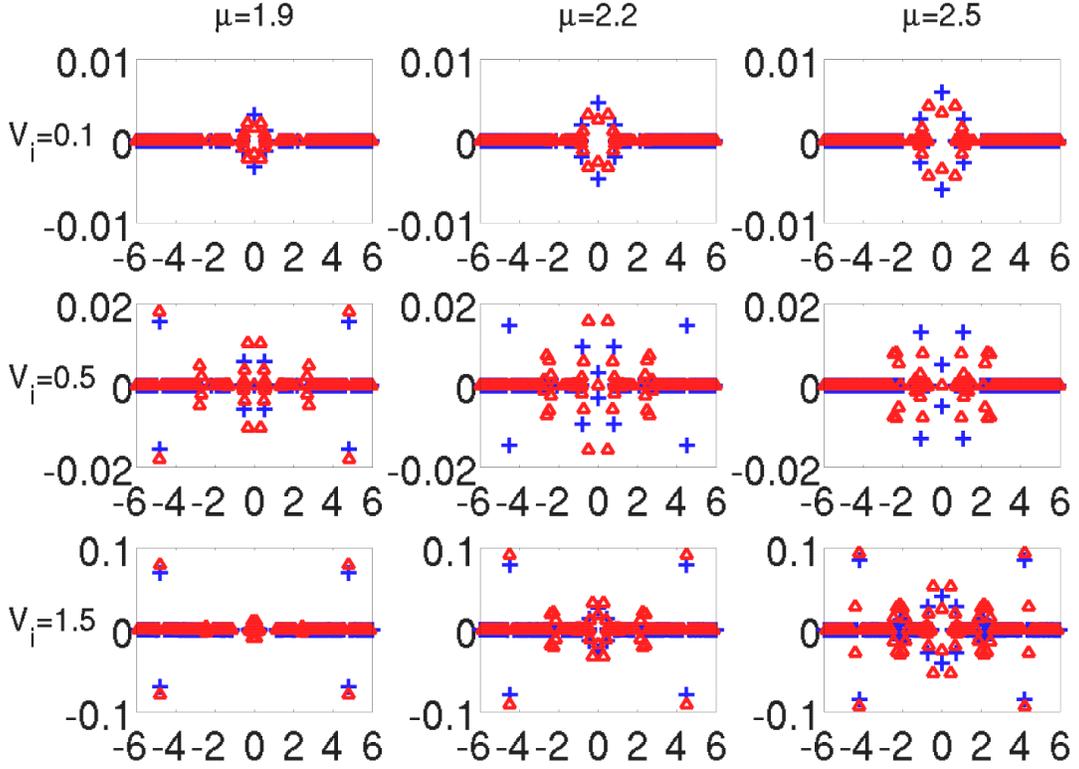} 
\caption{\label{Fig:eigenvalues_map} Eigenvalues $\lambda_g$ for various $\mu$ (i.e., soliton power) and imaginary potential $V_i$ in the complex plane (the horizontal axis corresponds to the real part, the vertical axis to the imaginary one). Blue $+$ and red $\triangle$  correspond to $\sigma=1$ and $\sigma=10$, respectively. }
\end{figure*}

The stability of the calculated gap soliton can be addressed by considering the effect of small perturbations in the form 

\begin{align}
\Psi(x,z) &=[\phi(x) + p(x)e^{i\lambda_g z} + q(x)e^{-i\lambda_g^* z}] e^{-i \mu z}, \label{eq:pert_psi} \\
V_\mathrm{NL}(x,z) &= V^\mu_\mathrm{NL}(x) + \Delta V_\mathrm{NL}(x,z), \label{eq:pert_potential}
\end{align} 
 
where $V^\mu_\mathrm{NL}$ is the nonlinear potential computed via Eq. \eqref{eq:sol_VNL}, corresponding to the soliton $\phi(x)$ once the propagation constant $\mu$, the degree of nonlocality $\sigma$ and the linear properties (i.e., $V_r$ and $V_i$) of the structure are selected. In the perturbative limit, i.e., neglecting nonlinear terms stemming from the added perturbation, we obtain from Eqs. (\ref{eq:phi})-(\ref{eq:sol_VNL}) the following linear eigenvalue problem 

\begin{eqnarray}
&& \lambda_g p= [\bm{L} -iV_i(x)]p -\phi \bm{D}\left(\phi^* p + \phi q^* \right) ,   \label{eq:lsa}\\
&& \lambda_g q^*= [-\bm{L} - iV_i(x)] q^* +\phi^* \bm{D}\left(\phi q^* + \phi^* p \right)  ,   \label{eq:lsa2}
\end{eqnarray}
where we defined the operator $\bm{L} = 0.5 {\partial_x^2} - V^{\mu}_\mathrm{NL}(x) - V_r(x) + \mu$. Moreover, we introduced  $\bm{D} = (1-\sigma{\partial_x^2})^{-1}$, capable to compute the field intensity  from the nonlinear potential $V_\mathrm{NL}$; in other words, $\bm{D}$ is the convolution between the Green function of Eq. \eqref{eq:Yukawa} and the intensity profile. \\
\begin{figure}
\centering
\includegraphics[width=0.49\textwidth, height=5.5cm]{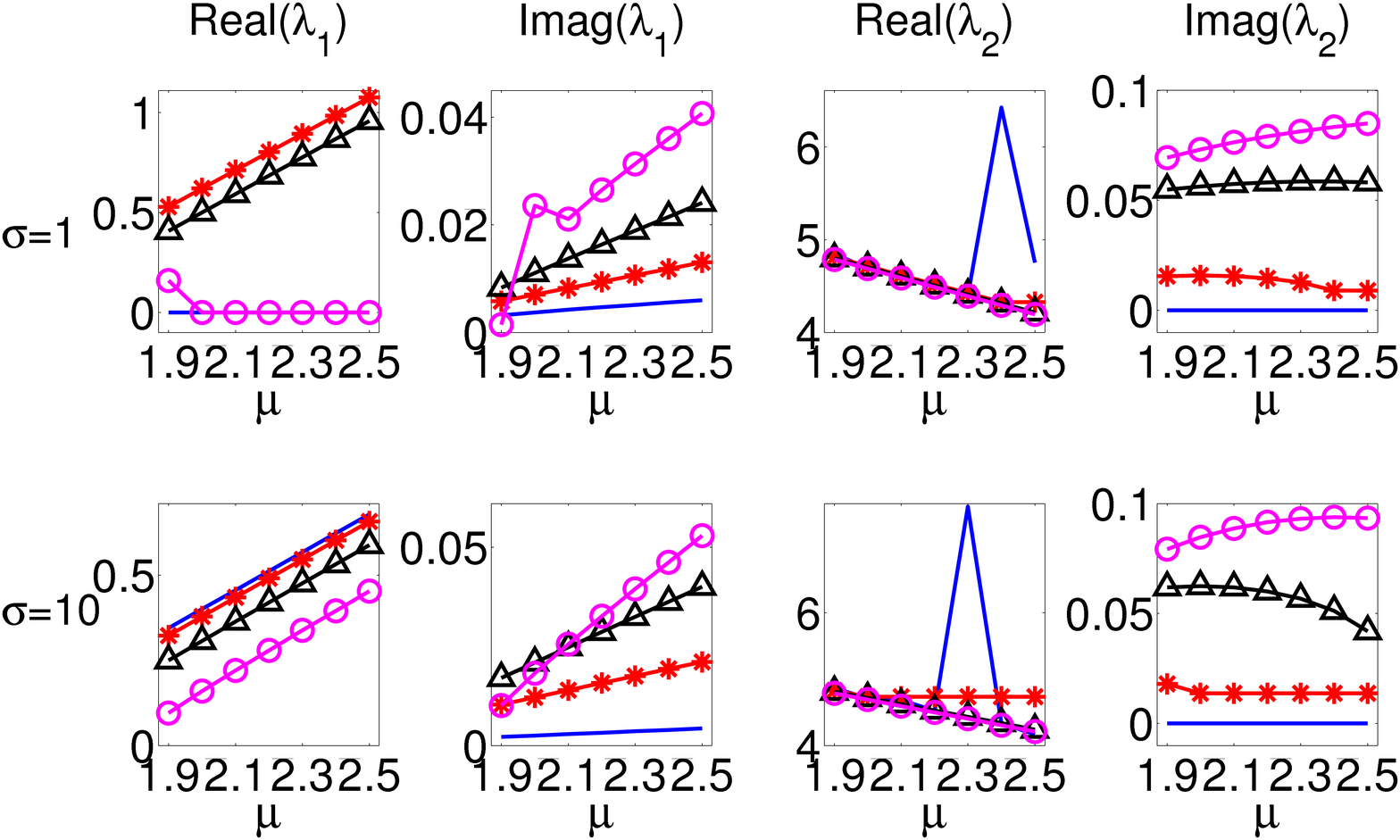}
\caption{\label{Fig:eigenvalues_max} $\lambda_1$ and $\lambda_2$ versus $\mu$, for $\sigma=1$ (first row) and $\sigma=10$ (second row). The imaginary potential $V_i$ is $0.1$ (blue lines without symbols), $0.5$ (red lines with stars), $1.0$ (black lines with triangles) and $1.5$ (magenta lines with circles), respectively.}
\end{figure}
The solution is stable if $\text{Im}(\lambda_g) = 0$ holds for all the eigenvalues, i.e., if the system has only real eigenvalues. The existence of complex eigenvalues corresponds to an oscillatory instability with $\text{Im}(\lambda_g) > 0$ ($<0$), the latter implying exponentially decaying (growing) modes together with intensity oscillations while the wave evolves along $z$. We solved the system of Eqs. (\ref{eq:lsa}-\ref{eq:lsa2}) using pseudo-spectral techniques based on Chebyshev polynomials to compute both the diffraction operator and the operator $\bm{D}$. We chose a grid extending to $40$ along $x$ (much larger than the maximum degree of nonlocality used, i.e. $\sigma=10$) to avoid artifacts. The  grid consisted of 1001 points, the latter ensuring independence from the numerical resolution  for both the eigenvalues and the eigenfunctions in the range of interest (see below). 

\begin{figure*}
\centering
\includegraphics[width=0.8\textwidth]{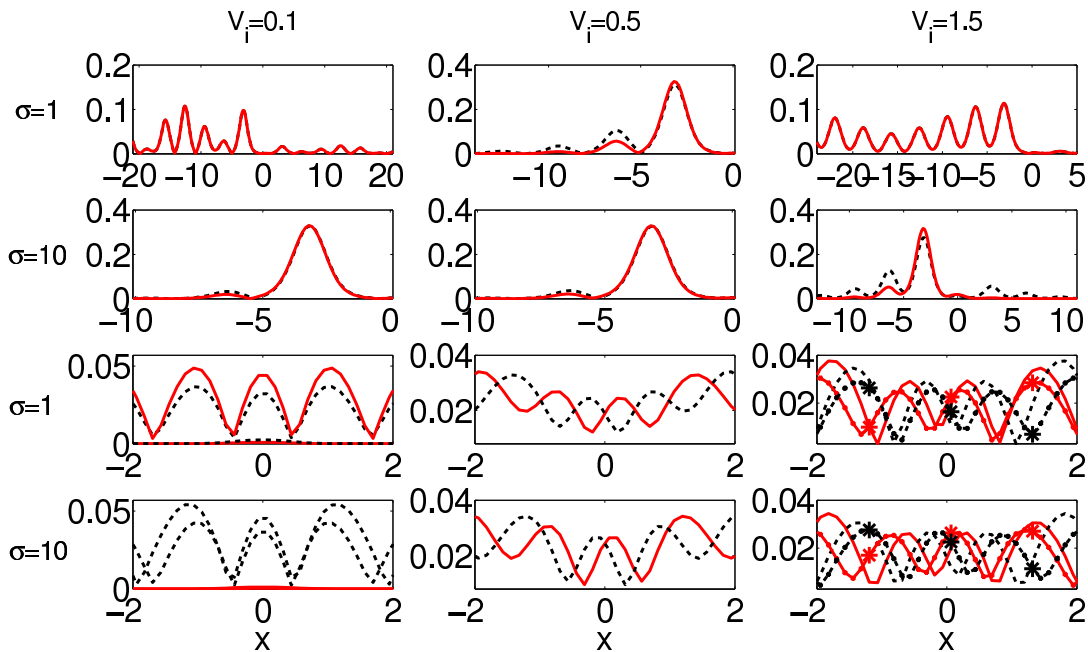} 
\caption{\label{Fig:eigenfunction} Spatial profile versus $x$ of the eigenfunctions corresponding to the eigenvalues  in Fig. \ref{Fig:eigenvalues_max}. The first two rows correspond to $\lambda_1$, the other two rows to $\lambda_2$. In the first two rows, red solid and black dashed lines correspond to $|p(x)|$ ($q(x)$ is negligibly small compared with $p(x)$) for $\mu=1.9$ and $\mu=2.5$, respectively; in the plots we took only the modes with $\text{Re}(\lambda_1)>0$ and $\text{Im}(\lambda_1)<0$. In the third and fourth rows, for $V_i=0.1$ red solid and black dashed lines correspond to $|p(x)|$ and $|q(x)|$, respectively; for larger $V_i$ the red solid lines with and without symbols correspond to $|q(x)|$ for stable perturbations ($|p(x)|$ is not plotted as it has the same behavior), whereas the black dashed lines with and without symbols correspond to growing modes. Lines without and with symbols correspond to $\mu=1.9$ and $\mu=2.5$, respectively. }
\end{figure*}
As a first attempt, we solved system (\ref{eq:lsa})-(\ref{eq:lsa2}) for $V_i=0$ and any degree of nonlocality: in agreement with previous works \cite{Jisha:2010}, gap solitons are stable (that is, $|\text{Im}(\lambda_g)|$ is less than $10^{-13}$, well below our numerical accuracy) for every $\sigma$ and $\mu$. Next, we considered $V_i\neq 0$. In the Kerr case ($\sigma$=0) we find solitons are unstable, as briefly described in Ref. \cite{Nixon:2012}; here we want to address the role played by nonlocality in soliton stability. Figure \ref{Fig:eigenvalues_map} illustrates the behavior of $\lambda_g$ versus the gain/loss coefficient $V_i$ and the soliton propagation constant $\mu$. The values of $\text{Re}(\lambda_g)$ are limited to the interval $[-6\ 6]$, as values outside it are associated with high frequency noise. First,  regardless of the soliton and structure parameters, there is an oscillatory instability due to the ubiquitous presence of eigenvalues with non-vanishing real as well as imaginary parts. Moreover, 
all 
the eigenvalues responsible for  the OI appears in quartets, featuring  the same  $|\text{Im}(\lambda_g)|$ and $|\text{Re}(\lambda_g)|$ \cite{Kevrekidis:2003}. The eigenvalue  distribution  is quite complex, nevertheless some general trends can be  observed. For $V_i=0.1$ the OI eigenvalues are located close to the origin of the complex plane, whereas for higher $V_i$ new branches of OI eigenvalues appear, stemming from the broken degeneracy of the purely real eigenvalues. We also note that a new quartet of OI eigenvalues appears for $V_i\geq0.5$, with $|\text{Re}(\lambda_g)|$ close to $4$.

The trend  of the instability versus $V_i$ and $\mu$ can be assessed by looking at Fig. \ref{Fig:eigenvalues_max}. To completely study the instability we computed the eigenvalues having the largest imaginary part (i.e., the maximum growth rate) in the range $|\text{Re}(\lambda_g)|\in [0\ 3]$ (we name it $\lambda_1$) and in the interval $|\text{Re}(\lambda_g)|\in [3\ 8]$ (we name it $\lambda_2$). Let us start from $\lambda_1$: in general, the growth rate increases with $\mu$, and at the same time $|\text{Re}(\lambda_1)|$ gets larger, that is, the leading unstable mode moves away from the origin. The former statement always holds true  except for $\sigma=1$ and $V_i=0.1$ or $V_i=1.5$: in both these cases the real part of $\lambda_1$ vanishes, i.e., there is no OI. By looking at the growth rate, for $V_i=1.5$ we see a rapid noise amplification, whereas for $V_i=0.1$ $|\text{Im}(\lambda_1)|$ is negligible, that is, the instability should be appreciable only over lengths much longer than the beam
Rayleigh distance. Finally, we also note that the growth rate  tends to increases with $V_i$.\\
The behavior of $\lambda_2$ is different. For low $V_i$ the growth rate is zero for every nonlocality $\sigma$. The real part of $\lambda_2$ does not depend on $V_i$, whereas it drops off as $\mu$ increases. Interestingly, the growth rate associated with $\lambda_2$ is dominant with respect to $\lambda_1$ for large $V_i$ (in Fig. \ref{Fig:eigenvalues_max} this occurs  for both $V_i=1$ and $V_i=1.5$).

We can summarize our findings as follows: except for the cases discussed above, the instability rate grows  with both $V_i$ and $\mu$. This relates to a break up of the $\mathcal{PT}$-symmetry when a perturbation is added to the soliton, analogously to Ref. \cite{Makris:2008}. In fact, large $V_i$ induce the appearance of a complex spectrum, see Fig. \ref{Fig:band} for example; in a similar way,  large $\mu$ correspond to lower real potentials trapping the wave, thus leading to a reduced effective $\mathcal{PT}$-breaking threshold.\\
To validate our interpretation of soliton stability, Fig. \ref{Fig:eigenfunction} graphs the eigenfunctions corresponding to the eigenvalues $\lambda_1$ and $\lambda_2$. For $\lambda_1$ the eigenfunctions $p(x)$ and $q(x)$ are strongly asymmetric with respect to $x=0$: the eigenfunctions featuring a positive growth rate (i.e., $\text{Im}(\lambda_1)<0$) are centered around $x=-\pi$, i.e., on the adjacent channel with respect to the gap soliton; eigenfunctions featuring a negative growth rate corresponding to an exponential decay on propagation (i.e., for $\text{Im}(\lambda_1)>0$, not shown in Fig. \ref{Fig:eigenfunction}) can be found by a mirror reflection with respect to $x=0$. Moreover, the transverse phase profiles of $p(x)$ and $q(x)$ are non uniform. All these properties confirm that the instability is due to the presence of perturbation modes breaking the $\mathcal{PT}$ symmetry: otherwise stated, the gap soliton  propagates in a \textit{sea} of unstable modes, excited by differences between the 
actual field profile and the exact soliton shape. The LSA allows us to predict that the soliton instability takes place as an asymmetric transfer of power towards the gain regions (in our geometry $x<0$), manifesting itself together with longitudinal oscillations in the field intensity due to the non-vanishing real part of $\lambda_1$.\\
For $\lambda_2$ the eigenfunctions are more complicated: for $V_i>0.1$ they appear as delocalized Bloch waves (in Fig. \ref{Fig:eigenfunction} a zoom around $x=0$ is plotted) with spatial frequency dictated by $\text{Re}(\lambda_2)$ and conserving the main properties found for $\lambda_1$: asymmetry around $x=0$, positive and negative growth rates, a non-flat phase profile. A difference exists due to their periodic profile along $x$: their shape is such that it is impossible to easily determine the drift direction, or its occurrence at all.

\subsection{Evolution of non-soliton solutions}
\label{sec:evolution_vs_power}

\begin{figure}[h!]
\centering
\includegraphics[width=0.49\textwidth, height=6cm]{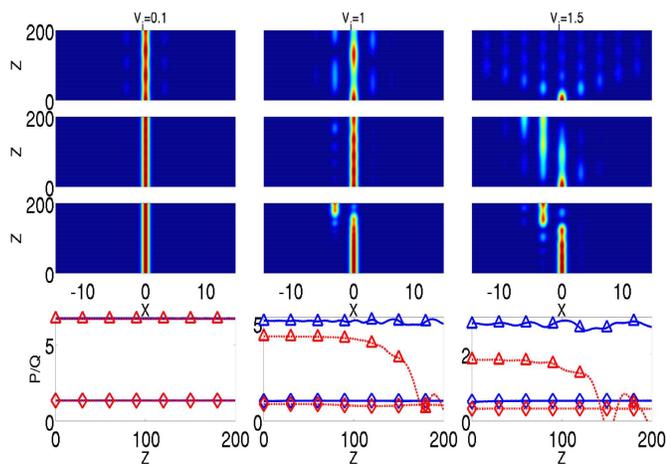}
\caption{\label{Fig:diff} Intensity evolution in the plane $xz$ when the shape of the input beam corresponds to the gap soliton, but with $0.25$ of the soliton power (first row), $0.5$ of the soliton power (second row) and  $1.25$ times the soliton power (third row); here we chose $\mu =2$ and $\sigma=10$. The last row illustrates the $z$ evolution  of parameters $P$ (blue solid line) and $Q$ (red dotted line) for powers $0.25$ ( $\diamond$) and $1.25$ ($\triangle$ ).  }
\end{figure}

\begin{figure}
\centering
\includegraphics[width=0.5\textwidth, height=6.1cm]{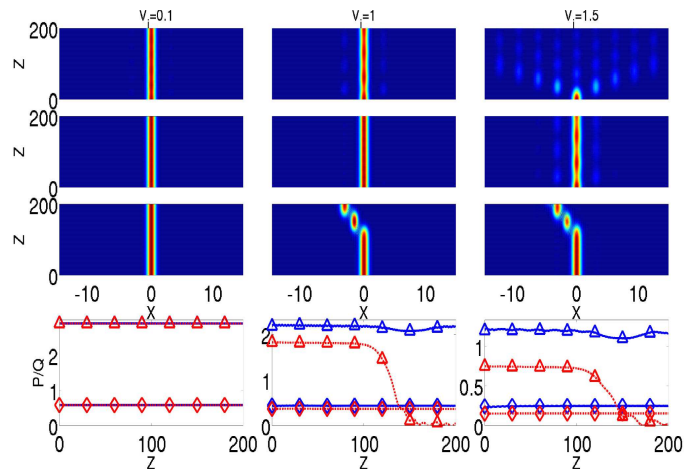}
\caption{\label{Fig:diffd1} Same as in Fig. \ref{Fig:diff} but with $\sigma=1$.  }
\end{figure}

Before proceeding with the dynamical stability analysis of the stationary solution, we investigate the trade-offs between diffraction, linear lattice and self-defocusing nonlinearity on wave propagation by numerically integrating Eq. \eqref{eq:GPE} with a standard Beam Propagation Method (BPM); we use splitting of the propagation operator and a Crank-Nicolson scheme for the diffraction term. We launch in the nonlinear lattice a wavepacket with spatial distribution corresponding to a gap soliton, but varying its input amplitude. This approach corresponds to adding low frequency noise to the soliton. We start with an input excitation which is one quarter of  the soliton power; Fig. \ref{Fig:diff} (top row) plots the corresponding numerical results. Discrete diffraction is observed, with the linear $\mathcal{PT}$-symmetric potential introducing a left/right asymmetry in the intensity distribution owing to the non-reciprocity of the Bloch-Floquet 
modes \cite{Makris:2008}. 

Accordingly, the asymmetry becomes prominent as the imaginary part of the refractive index $V_i$ gets larger (compare different columns in Fig. \ref{Fig:diff}). The complex potential also results in longitudinal oscillations of the beam intensity versus propagation, in agreement with the presence of OI. A further increase in power to half that of the soliton  reduces diffraction, but for increasing $V_i$ the symmetry in propagation is broken and the beam diffracts  only in one direction (second row of Fig. \ref{Fig:diff}). Increments in excitation reduce discrete diffraction until, when the power corresponds to the exact soliton, the wavepacket profile along $z$ becomes invariant; noteworthy, in agreement with Fig. \ref{Fig:profiled}, the larger $V_i$ is the larger is the amount of power coupled to adjacent guides owing to a lower linear confinement.

For powers above soliton generation (Fig. \ref{Fig:diff}, third row) self-defocusing becomes comparable with the linear trapping potential $V_r$, until eventually  a power-dependent breaking of $\mathcal{P}\mathcal{T}$-symmetry (see Fig. \ref{Fig:band}) occurs and, consequently, a  strong left/right asymmetry \cite{Makris:2008}. The net effect is a transverse motion of the nonlinear wavepacket from one channel to the other,  the direction dictated by $V_i$: the wave is attracted towards the gain region (in our case (see section \ref{sec:ngs}) negative $x$): this is in perfect agreement with the LSA carried out in Sec. \ref{sec:LSA}.

Even though the LSA is formally valid only for small perturbations, its results can describe other minor features of the wave evolution: the oscillation period increases with $V_i$ (LSA predicts a period of about $60$ for $V_i=1.5$, in agreement with BPM simulations) and 
the exponential growth gets larger with $V_i$ as well (for $V_i=1.5$ LSA predicts an increase equal to $e$ over a length of 50, in  agreement with BPM); the LSA can also account for the small excitation of the adjacent guide on the right side (i.e., towards the loss region) for large $V_i$ (see the third panel in the second row of Fig. \ref{Fig:eigenfunction}).  \\ 
Finally, Fig. \ref{Fig:diff} graphs the quantity $Q$, defined as the quasi-power $Q= \int \Psi(x,z) \Psi^*(-x,z) dx $ which is conserved  in the linear regime \cite{Makris:2008} together with the real power $P$: the power is almost conserved everywhere whereas $Q$ changes strongly when the $\mathcal{P}\mathcal{T}$-symmetry is broken due to the nonlinear response.\\
It is also important to understand the role played by nonlocality. To this extent, we repeated the simulations  in Fig. \ref{Fig:diff} for a lower nonlocality, in particular for $\sigma=1$, as displayed in Fig. \ref{Fig:diffd1}: in agreement with the LSA, in this case the gap soliton is slightly more stable, with a trend opposite to that of solitons for $V_i=0$, both in the discrete \cite{Lederer:2008} and in the continuous cases \cite{Snyder:1997}.

\begin{figure*}
\centering
\includegraphics[width=0.89\textwidth]{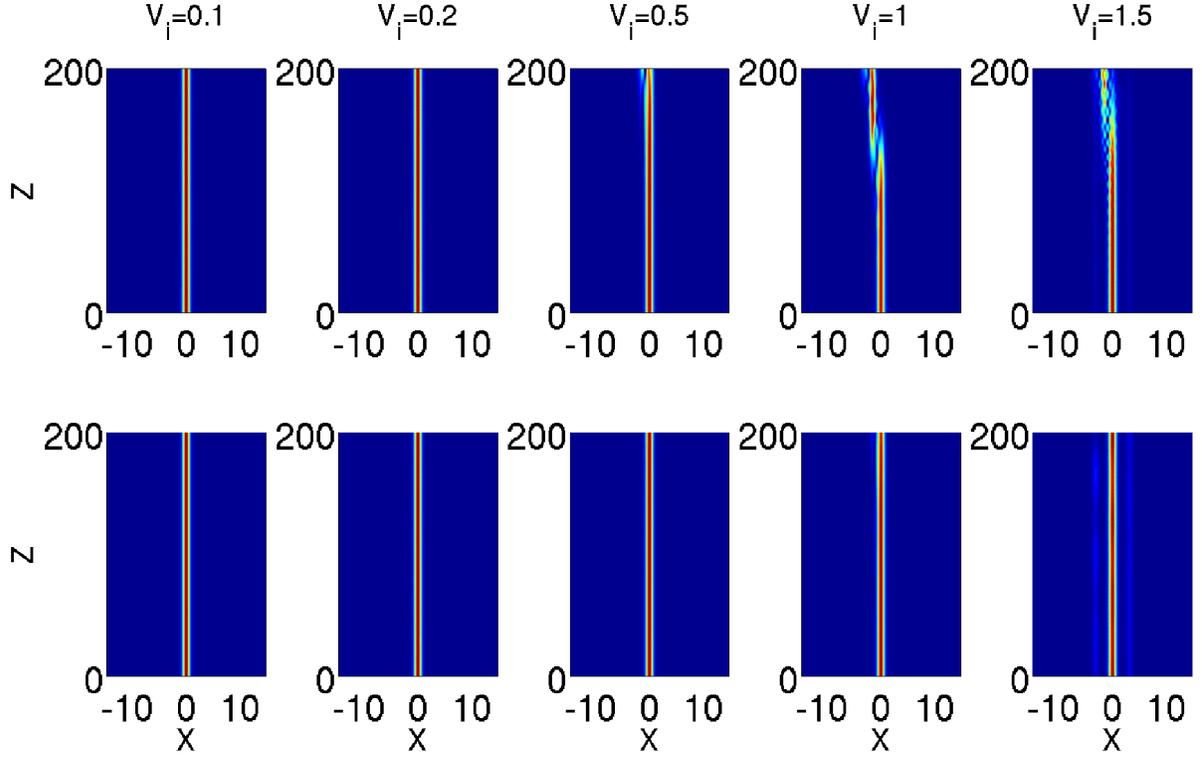}
\caption{\label{Fig:evolution1} Evolution of stationary solutions in the plane $(x,z)$ for $\mu = 1.9$. The soliton was heavily perturbed  by noise with standard deviation $0.01$ added to the input for $\sigma=1$ (top) and $\sigma=10$ (bottom), respectively, for various $V_i$ as marked. }
\end{figure*}

The wavepacket behavior versus  input power can be explained by resorting to particle conservation, as expressed by Eq. \eqref{eq:particle_conservation}. After recasting the divergence of the flux $j$ as $( \partial \rho/\partial x) \partial \chi/ \partial x + \rho \partial^2 \chi /\partial x^2$, the particle conservation expressed by Eq. \eqref{eq:particle_conservation} provides  $\partial \rho /\partial z>0$ when 

\begin{equation} \label{eq:condition_rho}
\frac{\partial^2 \chi}{\partial x^2} +  \frac{1}{\rho} \frac{\partial \rho}{\partial x} \frac{\partial \chi}{ \partial x} < 2V_i(x).
\end{equation}

According to Fig. \ref{Fig:profiled}, corresponding to the soliton we can set $\chi(x)\approx c_0 x$, i.e., the phase follows a linear trend across $x$, with $c_0$ a constant proportional to $V_i$; from the figure we find $c_0>0$. Equation \eqref{eq:condition_rho} turns into $(c_0/\rho) \partial \rho/\partial x<2V_i(x)$. First, we note that the amplitude change cancels out due to the linearity of the conservation equation in $\rho$. Second, due to self-defocusing nonlinearity, powers exceeding the soliton case yield a broadening of the soliton intensity profile $\rho$ (see Fig. \ref{Fig:var}), with a diminished absolute value of the derivative of $\rho$. This means that condition \eqref{eq:condition_rho} is satisfied for $x<0$ ( gain region), whereas it is broken for $x>0$ (loss region).
In other words, particles undergo accumulation in the gain region and depletion in the loss region due to the imbalance of the flux $j$. 

Let us now consider how this affects power coupling between adjacent guides. The accumulation of particles on the left of the core guide increases the net number of particles tunneling to the next channel, this being enhanced by  the defocusing character of the nonlinearity, i.e., by lowering of the Peierls-Nabarro barrier; conversely, in self-focusing media the particle flux outwards is reduced by the nonlinearity; hence, gap solitons are stable \cite{Musslimani:2008}. The opposite phenomenon takes place on the right side, with particles moving to the core guide from the lateral. Summarizing, the net effect is a particle motion towards negative $x$. Noteworthy, the field increase towards the lateral guide progressively reduces the nonlinear dephasing between them, allowing a partial back-coupling of power towards the input guide and thus inducing the oscillatory instability. The simulations demonstrate that, over several cycles, the particles tend to acquire a net motion towards negative $x$. Analogous 
phenomena occur for powers below soliton formation, but with beam broadening due to diffractive spreading.\\ 
The explanation just provided agrees  with both LSA and BPM results: the instability of gap solitons is related with the transverse flux of particle, thus it is very small for low $V_i$ (gap solitons for $V_i=0$ are stable) whereas is enhanced as the size of the gain/loss ratio is bigger. Additionally, flux considerations allow us to understand the dependence of soliton stability on nonlocality: a higher nonlocality yields narrower solitons, that is, larger $\partial \rho/\partial x$, in turn increasing the flux $j$. 

\begin{figure*}
\centering
\includegraphics[width=0.89\textwidth]{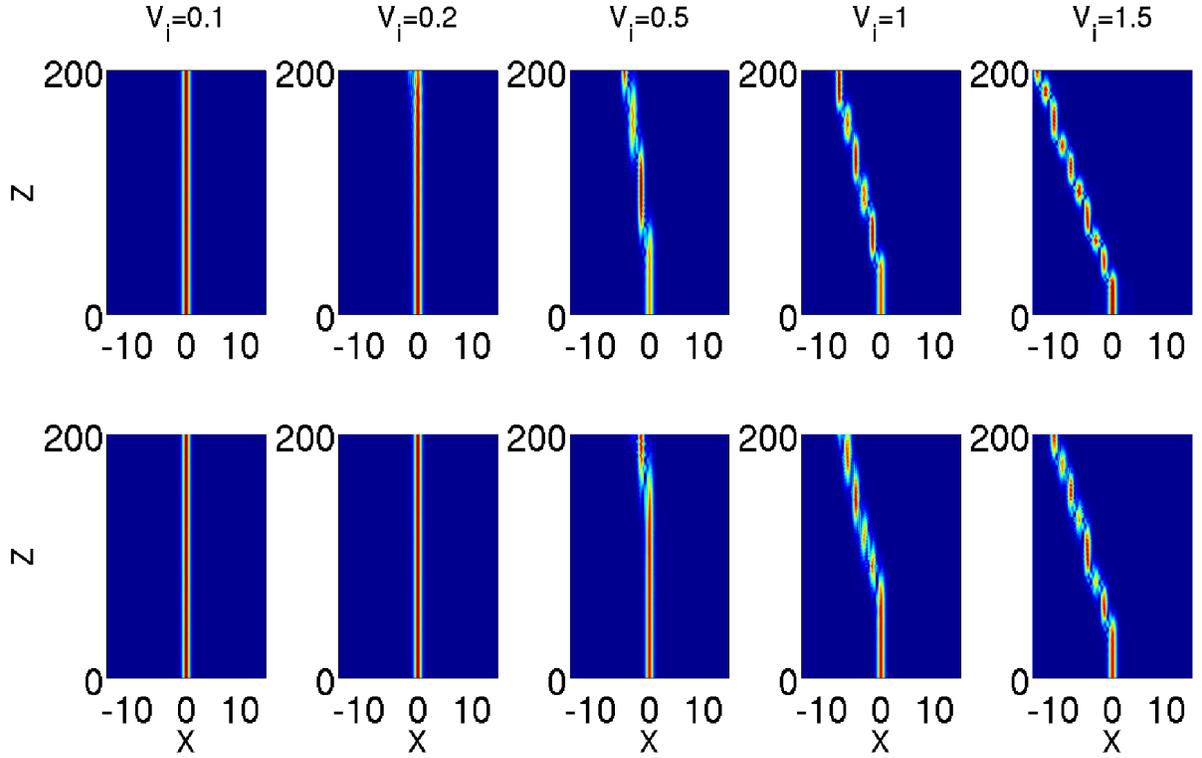}
\caption{\label{Fig:evolution2} Same as in Fig. \ref{Fig:evolution1} but for $\mu = 3$. }
\end{figure*}

\subsection{BPM study of noise effects on soliton propagation}

Next we analyze the dynamical stability of propagating gap solitons by adding Gaussian noise at the input section $z=0$ with a magnitude corresponding to the $0.01\%$ of the soliton amplitude. 
For low $V_i$  LSA predicts $|\text{Im}(\lambda_g)|$ on the order of $10^{-3}$, that is, the instability should appear for propagation lengths larger than $10^3$ (blue solid lines in Fig.~\ref{Fig:eigenvalues_max}) (for the sake of comparison with the linear regime, discrete diffraction induces an appreciable spreading at $z=20$). 
The numerical simulations do not show appreciable changes in soliton profile up to $z=200$, thus confirming the LSA results, regardless of the value of $\mu$ (first two columns in Figs.~\ref{Fig:evolution1} -  \ref{Fig:evolution2}). For larger $V_i$  LSA predicts a much higher growth rate, comprising oscillations in the intensity evolution versus $z$ and growing modes for all $\sigma$:  the numerical results in the last two columns of Figs. ~\ref{Fig:evolution1}-~\ref{Fig:evolution2} confirm the predictions. Consistently with the LSA, the simulations demonstrate that, for $\mu$ closer to the edge of the bottom band, the soliton stability improves more than for $\mu$ closer to the edge of the top band, i.e., the growth rate increases with $\mu$. \\
The LSA predicts similar instability lengths for solutions with different $\sigma$ and large $V_i$, with instability in general increasing as nonlocality becomes larger, the latter finding being confirmed in presence of low-frequency noise, see Sec. \ref{sec:evolution_vs_power}. Conversely, BPM simulations shown in Figs. ~\ref{Fig:evolution1}-~\ref{Fig:evolution2}  indicate that the instability drops off when nonlocality is increased, but confirming the small dependence of the growth rate from nonlocality. The observed behavior can be explained in the context of LSA: when we add high-frequency noise, we are exciting unstable modes with large $\text{Re}(\lambda_g)$ (larger than 8, thus out of the range plotted in Fig. \ref{Fig:eigenvalues_map}), encompassing a complicated distribution of the eigenvalues versus nonlocality. The last statement is confirmed by the high-frequency variations in the intensity distribution in Figs. ~\ref{Fig:evolution1}-~\ref{Fig:evolution2} in comparison with the smooth behavior 
followed in Figs. \ref{Fig:diff}-\ref{Fig:diffd1}.

\begin{figure}
\centering
\includegraphics[width=0.49\textwidth]{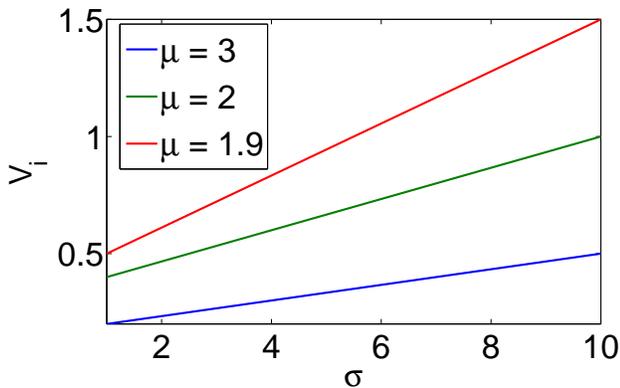}
\caption{\label{Fig:instability} Magnitude of the imaginary potential $V_i$ at which instability occurs for various nonlocality strengths  $\sigma$ and  increasing $\mu$ (top to bottom curves). Solitons remain stable for a larger range of $V_i$ when  $\mu$ is lower. } 
\end{figure}

Finally, Figure \ref{Fig:instability} provides an estimation of the instability behavior plotting the value of $V_i$ at which the instability manifests on a distance lower than 200, when a soliton perturbed with a noise of constant amplitude (0.01), regardless of the nonlocality parameter $\sigma$ and of the propagation constant $\mu$. As discussed previously, instability increases both with $V_i$ and $\mu$; at the same time, broader nonlinear response helps in inhibiting soliton blow-up.\\

\section{Conclusions}

In conclusion, we investigated shape, existence curve and stability of $\mathcal{PT}$-symmetric gap solitons in a self-defocusing medium featuring  a nonlocal nonlinearity and a linear periodic  potential. We showed that fundamental gap solitons exist inside all the first bandgap, with appreciable tails in proximity of the top of the lowest band edge.
 As in the absence of gain/loss, the magnitude of the tails decreases as the nonlocal range broadens; moreover, for a given propagation constant, a higher nonlocality requires higher powers to form a soliton. We also demonstrated that solitons possess a real and imaginary parts in order to conserve particle number, with the imaginary part proportional to gain/loss terms. A variational approach suitable for the study of $\mathcal{PT}$-symmetric solutions was developed, as well as an analytical method in the highly nonlocal limit, confirming the same dependence of soliton features on system parameters as showed by exact numerical solutions. We found that, in the presence of an imaginary potential, gap  solitons become oscillatory unstable. Moreover, using both a linear stability analysis and BPM simulations, we showed that the perturbation growth rate  changes dramatically with the imaginary potential: for large gain/loss terms the soliton shape is conserved over much shorter distances than in the case of 
small imaginary potentials (much shorter than the characteristic discrete diffraction length). The instability manifests mainly as a transverse particle flux (photons in the electromagnetic case) across the periodic lattice, with soliton motion towards the gain region. We also demonstrated that instability slightly changes with the response width of the nonlinearity, with the behavior being strongly dependent on the spectral contents of the applied perturbation. Finally, we demonstrated that solitons closer to the lower band-edge, i.e., with lower $\mu$, are more stable than those with high $\mu$, the former solitons being  well within the region where $\mathcal{PT}$-symmetry is fulfilled. Last, our findings can find application, for example, in the power-driven control of optical signals into a waveguide array.

\section*{Acknowledgements}
JCP gratefully acknowledges FCT grant n. SFRH/BPD/77524/2011 for support and thanks YuanYao Lin for useful discussion.


%


\end{document}